# ALOHA With Collision Resolution(ALOHA-CR): Theory and Software Defined Radio Implementation

Xin Liu, John Kountouriotis, Athina P. Petropulu and Kapil R. Dandekar

Electrical & Computer Engineering Department, Drexel University, Philadelphia PA

[1]

### Abstract

A cross-layer scheme, namely ALOHA With Collision Resolution (ALOHA-CR), is proposed for high throughput wireless communications in a cellular scenario. Transmissions occur in a time-slotted ALOHA-type fashion but with an important difference: simultaneous transmissions of two users can be successful. If more than two users transmit in the same slot the collision cannot be resolved and retransmission is required. If only one user transmits, the transmitted packet is recovered with some probability, depending on the state of the channel. If two users transmit the collision is resolved and the packets are recovered by first over-sampling the collision signal and then exploiting independent information about the two users that is contained in the signal polyphase components. The ALOHA-CR throughput is derived under the infinite backlog assumption and also under the assumption of finite backlog. The contention probability is determined under these two assumptions in order to maximize the network throughput and maintain stability. Queuing delay analysis for network users is also conducted. The performance of ALOHA-CR is demonstrated on the Wireless Open Access Research Platform (WARP) test-bed containing five software defined radio nodes. Analysis and test-bed results indicate that ALOHA-CR leads to significant increase in throughput and reduction of service delays.

*keywords*-**multi-user system, blind source separation, MIMO systems, collision resolution, software defined radio**

## I. INTRODUCTION

In a wireless uplink scenario, collisions occur when two or more users transmit at the same time over the same channel. Traditionally, once a collision occurs, retransmissions are requested. Retransmissions lower throughput and waste power and bandwidth. Well studied schemes for avoiding collisions include Carrier Sensing Multiple Access with Collision Avoidance (CSMA/CA) (e.g., see IEEE 802.11 [1]). In order to overcome the hidden terminal problem, IEEE 802.11 incorporates a positive acknowledgment scheme, i.e., Request To Send (RTS) followed by Clear To Send (CTS). However, in most protocols, collisions occur more frequently as the traffic load increases, in which case the RTS/CTS scheme becomes less effective due to collisions of the RTS reservation packets.

[1]This work has been supported by the Office of Naval Research under Grant ONR-N-00014-07-1-0500.

Collision resolution can be viewed as multiuser separation. However, well known approaches that allow for multiuser separation, such as TDMA, FDMA, OFDMA, CDMA or use of multiple antennas, might not be a good fit for wireless networks. Wireless network traffic can be bursty, users operate on limited battery power, and in certain cases, wireless receivers have physical size limitations. TDMA, FDMA and OFDMA approaches are fixed resource allocation schemes and thus are not efficient for bursty traffic. The CDMA approach requires bandwidth expansion, which results in increased power consumption for each wireless network user. The use of multiple antennas, might not be feasible for compact wireless receivers. Wireless network-friendly approaches to achieve diversity include the NDMA protocol [6, 19], ALLIANCES [10, 17] and ZigZag decoding [7]. In these protocols, collisions are resolved by combining collided packets and several retransmissions. In these schemes it is assumed that nodes transmit with the same power, and that there is no significant power decrease due to propagation in small-scale networks. For cases in which users transmit using different power levels, user separation could be achieved via successive interference cancellation (SIC) [14]. However, it might not be a good approach to assign different power to different users. Collisions only happen with some probability, therefore, it would not be good for a user to transmit at low power all the time just to be separable in the event of a collision. A potentially network-friendly approach that does not require retransmissions and allows users to transmit at the same power was recently proposed in [12, 11, 18]. According to [12, 11, 18], by upsampling the received signal and viewing its polyphase components as independent linear mixtures of the collided packets, under certain conditions, the collided packets can be recovered in a blind fashion based on a single collision. In [18], user separation was enabled by different carrier frequency offsets (CFO) and user delays. In [12, 11], pulse-shape diversity was investigated as source of additional diversity in case user delays and CFOs are small.

In this paper, a novel cross-layer scheme is proposed for high throughput wireless communications in a cellular scenario. Transmissions occur in a time-slotted ALOHA-type fashion but with an important difference: simultaneous transmissions of two users can be successful. The wireless channel is assumed to be flat fading and constant over the duration of one time slot. A user $i$ with a non-empty queue transmits a packet with some probability $p$ in the beginning of each time slot, after waiting for a random time interval $\tau_i$. Each user embeds orthogonal pilots in its packet. In each slot, the base station (BS) determines the number of transmitting users. If there are more than two users, the packets are discarded and the users are asked to retransmit at a later time. If there is only one user present, its packet is recovered with some probability, depending on the state of the channel. If there are two users present, the users are separated and their packets are recovered by first oversampling and exploiting independent information about the two users that is contained in the polyphase components of the received signal. The properties of the user delays $\tau_i$'s are determined so that the probability of user separation is maximized. The system throughput is derived under the infinite backlog assumption, i.e., the network users always have data in their queues, and also under the assumption of finite backlog. Queuing delay analysis for network users is also conducted.

The performance of the proposed approach is demonstrated via simulations, and also via experiments conducted on a software defined radio (SDR) [2] testbed. This experimental wireless network consists of five nodes, i.e., one base station (BS) or access point, and four users, and was deployed in an indoor laboratory environment. The experimental results suggest that 70% of two order collisions can be resolved by the BS under realistic conditions, which results in higher throughput and lower service delays.

*1) Relation to other published work:* Collision resolution is based on the ideas of [18], where naturally occurring user delays and carrier frequency offsets were shown to provide diversity that enables blind user separation, i.e., separation without knowledge of the channel. However, in our experimental setup, we observed that naturally occurring delays and CFOs are rather small. Thus, in this work, we ignore CFOs and introduce intentional delays in addition to the naturally occurring ones. The statistical characteristics of the intentional delays are chosen to enhance the separability of the users. In order to keep the complexity low and the probability of user separation high, resolution of only second order collisions is considered here. The work of [18] was concerned with the physical layer only. Here, we propose a cross-layer approach, and study throughput and queuing performance as well as physical layer issues. Further, a host of physical layer issues motivated by the real implementation are studied.

Multiuser separation based on user delays was also considered in [4]. The approach of [4] considers transmission of isolated frames; it exploits the edges of a frame over which users do not overlap, and assumes knowledge of the channel. However, noise can be a problem when exploiting edge effects as samples are taken at points where the pulse is quite low.

In relation to the collision resolution approaches NDMA [6, 19], ALLIANCES [10, 17] and ZigZag decoding [7], the proposed approach resolves collisions of order two without retransmissions. Thus, no storage of the collision signals is needed, and network users not involved in the collision do not need to wait until the collision is resolved.

*2) Organization:* The rest of the paper is organized as follows: The physical layer of ALOHA-CR is introduced in Section II and several implementation considerations are discussed in Section III. Network performance quantities for the Aloha-CR, like throughput and packet delays, are analytically derived in Section IV. A brief description of the SDR platform used to implement Aloha-CR appears in Section V, while specifics of our implementation of the physical and MAC layers on the SDR platform appear in section VI. The obtained experimental results are presented in Section VII and are compared with analytical and simulation results. Finally, conclusions are drawn in Section VIII.

*3) Notation:* Bold capitals denote matrices. Bold lower cases denote vectors. $H$ denotes transpose conjugate. The subscript $T$ denotes transpose. The subscript † denotes pseudo-inverse. $\| \cdot \|_F$ denotes Frobenius-norm. $\mathrm{D}iag(\mathbf{v})$ denotes the diagonal matrix with diagonal elements $v$. $\lceil \cdot \rceil$ denotes rounding up to the nearest integer.

## II. ALOHA-CR: Physical Layer

The channel between transmitter and receiver is assumed to be flat fading. Moreover, the channel is quasi-static, i.e., the channel remains unchanged over the duration of a packet.

If within a given time slot $K$ users transmit, the baseband signal received by the BS equals

$$y(t) = \sum_{k=1}^{K} a_k x_k(t - \tau_k) + w(t), \tag{1}$$

where $a_k$ denotes the channel coefficient between the $k - th$ user and the BS; $\tau_k$ is a random delay associated with the user $k$; $w(t)$ represents noise; and $x_k(t)$ is the $k$-th user signal, i.e., $x_k(t) = \sum_i s_k(i) p(t - iT_s)$ where $s_k(i)$ is the $i-$th symbol of user $k$; $T_s$ is the symbol interval; and $p(t)$ is a pulse shaping function with main lobe support $[-T_s, T_s]$. The mainlobe of neighboring pulses overlap by 50%.

In each symbol interval, the received signal is upsampled by a factor of $P$, with sampling locations at $t = iT_s + mT_s/P$, $m = 1, 2, \cdots, P$. The $m-$th polyphase component of the sampled output is

$$y_m(i) = \sum_{k=1}^{K} a_k h_{mk}(i) * s_k(i) + w_m(i) \tag{2}$$

where "$*$" denotes convolution, and $h_{mk}(i)$ equals $h_{mk}(i) = p(iT_s + \frac{mT_s}{P} - \tau_k)$ for $i = ... -2, -1, 0, 1, 2, ....$

Using a pulse with low sidelobe, at the sampling points over the $i-$th symbol interval, the only interference is from the $(i + 1)$-th symbol. Therefore, the channel $h_{mk}(i)$ can be approximated as of length 2. The IOTA pulse is a good choice for maintaining low intersymbol interference [9].

 

The convolutional MIMO problem can be transformed to a scalar one as

$$y_m(i) = \mathbf{h}_m \mathbf{D} \mathbf{s}(i) + w_m(i) \tag{3}$$

where $\mathbf{h}_m = [[h_{m1}(0) \ h_{m1}(-1)], ..., [h_{mK}(0) \ h_{mK}(-1)]]$ , $\mathbf{D} = \mathbf{diag}([a_1, a_1, \cdots a_K, a_K])$ and $\mathbf{s}(i) = \left[ \begin{array}{c} [s_1(i) \ s_1(i+1)] , \cdots , [s_K(i) \ s_K(i+1)] \end{array} \right]^T$.

Let us form the vector $\mathbf{y}(i)$ by appending $y_m(i)$, $m = 1, ..., P$. We have

$$\mathbf{y}(i) = \mathbf{A}\mathbf{s}(i) + \mathbf{w}(i); \tag{4}$$

where $\mathbf{A} = \mathbf{H}\mathbf{D}$ and $\mathbf{H} = [\mathbf{h}_1^T, \mathbf{h}_2^T, \cdots, \mathbf{h}_P^T,]^T$.

If pilot data are available, the matrix $\mathbf{A}$ can be estimated based on the pilot symbols, and then used for the recovery of the information bearing symbols. If no pilots are available, estimating $\mathbf{A}$ and then recovering $\mathbf{s}(i)$ is still possible by viewing (4) as a $P \times 2K$ instantaneous blind MIMO problem. Assuming that $P \geq 2K$, and under certain conditions on $\mathbf{A}$, the system is identifiable [18]. Any blind source separation algorithm (e.g., the JADE algorithm [5]) can be applied at this point to obtain an estimate of $\mathbf{A}$, i.e., $\hat{\mathbf{A}}$, within a column permutation ambiguity and a a constant diagonal matrix, $\mathbf{\Lambda}$, with complex nonzero diagonal elements, which represents phase ambiguity. These ambiguities are trivial, and are inherent in blind estimation problems. Based on $\hat{\mathbf{A}}$ and using a least-squares equalizer we can get the de-coupled signals $\hat{\mathbf{s}}(i)$, within permutation and phase ambiguities as

$$\hat{\mathbf{s}}(i) = (\hat{\mathbf{A}}^H \hat{\mathbf{A}})^{-1} \hat{\mathbf{A}}^H \mathbf{y}(i) = e^{jArg\{-\mathbf{\Lambda}\}} |\mathbf{\Lambda}|^{-1} \mathbf{P}^T \mathbf{s}(i) \tag{5}$$

where $\mathbf{\Lambda}$ is a diagonal matrix and $\mathbf{P}$ is a permutation matrix. Denoting by $\theta_k$ the $k-$th diagonal element of $Arg\{\mathbf{\Lambda}\}$, the $k-$th input signal can be recovered within a phase ambiguity as $\hat{s}_k(i) = s_k(i)e^{-j\theta_k}$. Although uniform sampling was described above, non-uniform sampling can also be done [18].

### A. About users' delays

For fixed sampling locations and a fixed pulse-shape function, the condition number of $\mathbf{A}$ can be controlled by the user delays $\tau_k$, where $k = 1, 2, ..., K$. If $\tau_k$'s are close to each other, the columns of $\mathbf{A}$ will be highly correlated, which results in high condition number.

Since naturally occurring delays are too small to guarantee a well conditioned $\mathbf{A}$, we propose that, before transmission, each node introduces an intentional delay. Let $\tau_k$ be the sum of the naturally occurring delay and the intentional delay. In this work we try to resolve collision of order two only. Let us express the delay difference between the two users is $\tau = \alpha + \delta$, where $\alpha$ is the difference between the intentional delays between users $i$ and $j$, and $\delta$ is the difference between the naturally occurring delays. Let $f_\delta(x)$ be the pdf of the natural delays differences, and further assume that $f_\delta(x)$ is symmetric around the origin.

*__Proposition 1__: Let the intentional delays be uniformly distributed over some interval $[0, T]$. If $T = T_s$, the probability of the collision being non-resolvable achieves a local minimum, independent of $f_\delta(x)$.*

The proof is given in Appendix A.

## III. Physical Layer Implementation Issues

Several issues need to be addressed in a practical implementation of the proposed approach.

### A. Frequency offsets and phase tracking

In a practical system there are always CFOs between transmitters and receiver, resulting from mismatch between transmitters and receiver oscillators, and also from Doppler shifts due to relative movement between transmitters and receiver. In this case the continuous-time base-band received signal $y(t)$ is of the form:

$$y(t) = \sum_{k=1}^{K} a_k x_k(t - \tau_k) e^{j2\pi F_k t} + w(t) \tag{6}$$

where $F_k$ is the CFO for user $k$. In [18] the CFOs were used as source of diversity that enables use separation. In the implementation that we consider here the CFOs are too small to provide diversity, and thus are ignored in the problem formulation. However, their effect is still present in the separated symbols, i.e.,

$$\hat{s}_k(i) = s_k(i) e^{j(-\theta_k + 2\pi F_k T_s i)} \tag{7}$$

The effect of the CFO on the separated signal can be mitigated using a phase locked loop (PLL) device [16]. The input output relationship for the PLL is $S_O(i) = S_I(i)e^{j\phi(i)}$, thus, the CFO estimate can be obtained as $\hat{F}_k = \frac{1}{2\pi} d\phi/dt$.

The phase ambiguity, $\theta_k$ can be compensated for through use of pilot symbols, or by using differential phase offset keying.

### B. Successive interference cancellation

Successive interference cancellation applied on a mixture of signals treats one of the components of the mixture as the signal of the interest and the rest as interference. The approach of Section II can be combined with SIC to further improve packet recovery performance. In particular, after blind source separation, the contribution of the strongest user signal can be reconstructed and deflated from the received signal. This usually provides a better estimate for the weak user.

One way to determine which is the *strongest user* is to look for the signal that has the smallest variance around the known constellation. Let the strong user be user $k$. Reconstruction of the contribution of the $k$-th user to the received signal requires knowledge of the pulse shape waveform, and estimates for CFO ($\hat{F}_k$), channel coefficient ($\hat{a}_k$), and delay ($\hat{\tau}_i$). The reconstructed signal is:

$$\hat{y}_k(t) = \hat{a}_k \sum_i \hat{s}_k(i) p(t - iT_s - \hat{\tau}_i) e^{j2\pi \hat{F}_k t}. \tag{8}$$

where $\hat{s}_k(i)$ are the estimated symbols.

The channel coefficient estimates can be obtained by cross-correlating the known pulse-shape waveform with the signal returned by the JADE algorithm. The user delays can be estimated at the synchronization step (see Section III-C), and the CFOs can be estimated as described in Section III-A.

Due to the delay between users, the peaks of different user pulses do not overlap. One could naturally wonder whether applying successive interference cancellation would be sufficient instead of upsampling the signal and performing packet separation as in Section II. As will be shown in Section VII via both simulation and testbed measurements, using SIC directly results in inferior results.

### C. Determining sampling points

In order to determine the beginning of the packet, some form of synchronization is required. For synchronization purposes, users are assigned distinct pseudo random sequences (pilots). The base station keeps a record of all pilot sequences in use in the network.

When the packet arrives, the base station uses the beginning part of the received signal to perform correlation with every entry of the code book. A peak in the correlation of the received signal with code $i$ indicates the presence of user $i$. The peak location provides the beginning of the packet of user $i$, while the peak value provides the corresponding channel coefficient.

This can be repeated for all possible users, however, in practice the following approach works better. The strongest user is identified as the one that produces the largest peak in the correlation. Then, the user pilot signal is reconstructed based on estimated channel coefficients and delays, and is subsequently deflated from the pilot portion of the received signal along the lines of Section III-B. The CFO effect is ignored here because of the short duration of the pilot segment.

For synchronization purposes, the best pulse shaping waveform for the pilots is the raised root cosine (RRC) function, as this function maximizes the SNR at the output of the matched filter [13] while it eliminates ISI at the sampling points.

We should note that the part of the packet containing the actual information will need to be oversampled in order for the method described in Section II to be applied. As explained in that section, the best pulse shaping for that purpose is the IOTA pulse [9]. However, we could not have used IOTA for the pilots, because the convolution of IOTA with itself introduces ISI at $t = nT_s$, thus the matched filter would not work well.

### D. Blind versus pilot-based user separation

Since a real communication system always uses pilots for synchronization purposes, one would think that these pilots could be used to estimate the matrix $\mathbf{A}$ in (4), which then could be used to recover the information bearing symbols. However, the fact that different pulse shape waveforms are used for pilots and information bearing symbols renders that approach impossible. As was already mentioned, in order to maximize the matched filter performance, RRC pulse shaping is used for the pilot symbols to be used in synchronization. Also, in order to minimize intersymbol interference between neighboring symbols of a user, IOTA pulse shaping is used for the payload symbols. Thus, the estimate of matrix $\mathbf{A}$ based on the pilots would be different than that corresponding to the payload (based on (3), $\mathbf{A}$ depends on the pulse shape function).

However, we can first estimate channel coefficients $a_k$ and user delay $\tau_k$ based on the pilot symbols, and subsequently combine them with the IOTA pulse shape function and sampling points (see (3)) to get the estimate of channel matrix $\hat{\mathbf{A}}$. Following the estimation of $\hat{\mathbf{A}}$ the symbols can be recovered via least-squares. We term this approach as *training method*. In VII we compared the training method to the blind approach, in which the matrix $\mathbf{A}$ is considered to be unknown. As it will be seen in that section, the estimation errors in channel coefficients and user delays render the training method inferior to the blind one.

### E. Dealing with collisions of any order and packet recovery

In theory, one could use the pilots to determine the number of users present in a collision. This is a detection problem and in low signal-to-noise ratio cases can lead to erroneous conclusions. Instead of attempting to estimate the collision order, we propose the following procedure.

The received signal is always treated as if it contained two users. The users are first separated as explained in Section II, and the strongest user is deflated from the received collision as described in Section III-B. For the remaining signal one of the following possibilities holds: (i) the signal is just noise; this is when there is only one user signal, or there are more than two users in the received signal. (ii) the remaining signal corresponds to the signal of the second user that was involved in the collision.

Although the above approach treats even the case of a single user as a potential collision, our experience indicates that this is a more robust approach than detecting the number of users first and then acting accordingly.

Following the synchronization step, regardless of the collision order, the incoming packet would be over-sampled by $4$. By applying the blind separation method of Section II we would get $4$ sequences. Each sequence would be passed through a PLL. Since the output of the PLL would be scattered around the nominal constellation the sequence with the smallest variance would be chosen. We refer to the strongest signal as $s_u(.)$. The decision on whether this is $s_u(i)$ or $s_u(i + 1)$ can be resolved using the user ID (i.e., the MAC address). Next, the strong user would be deflated from the received signal as discussed in Section III-B. If there was only one user in the received signal, or if there were more than two users, the remaining signal after deflation would not have a meaningful structure; this could be determined using the user ID. Otherwise, the deflation would yield the signal of the second user.

## IV. Throughput Analysis of ALOHA-CR

Consider a cellular network of $K$ users who communicate with a base station (BS). Users transmit their packets in a time slotted fashion with probability $p$. Each packet contains multiple symbols, and the time slot duration is equal to the packet duration plus two symbols.

The proposed ALOHA-CR schemes follows the slotted ALOHA protocol, expect that second-order collisions can be resolved. Since collision of packets can be resolved, it is expected that ALOHA-CR will have higher throughput than slotted ALOHA. In this section we first analyze the throughput of ALOHA-CR for the simple case of a network with infinite backlog, i.e. the case in which the queue of the nodes can never be empty, and each node always contends with some probability. In this case we analyze the throughput of the network. Then, we consider the case in which the nodes have finite backlog and analyze the throughput and service delay of ALOHA-CR.

The throughput is defined here as the number of successfully delivered packets per slot. We consider a network of $J$ users with $J > 2$, and each user contends with probability $p$. The following possibilities exist for each slot.

- No transmissions are attempted (empty slot).
- A single transmission is attempted. In this case, let depending on the channel state the probability of successful reception be $P_0$.

- Two transmissions are attempted. Let the probability of receiving correctly both of the transmissions be $P_1$ and the probability of successfully receiving only one of the two transmitting messages be $P_2$ (i.e., the probability of failing to receive any of the messages is $1 - P_1 - P_2$).

- More than two transmissions are attempted. In this case none of the transmitted messages can be successfully received and users have to retransmit at some later time.

### A. Network with infinite backlog and infinite number of users

For slotted ALOHA, the throughput is well established as $C(J) = Jp(1-p)^{J-1}$, which is maximized for $p^* = 1/J$ with maximum throughput $C(p^*) \to e^{-1}$ as $J \to \infty$.

For ALOHA-CR, the maximum throughput and optimum contention probability are given in the following proposition.

***Proposition 2***:

*The maximum network throughput is:*

$$C = \frac{2P_0^2}{P_0 - 2P' + \sqrt{P_0^2 + 4P'^2}} \left( 1 + \frac{2P'}{P_0 - 2P' + \sqrt{P_0^2 + 4P'^2}} \right) e^{\frac{-2P_0}{P_0 - 2P' + \sqrt{P_0^2 + 4P'^2}}}, \tag{9}$$

*where $P' = P_1 + \frac{P_2}{2}$, and is achieved for contention probability equal to:*

$$p = \frac{2P_0}{2P_0 + a + \sqrt{a^2 + 4P_0 b}} \tag{10}$$

*where $a = (P_0 - 2P')(J-1)$ and $b = P'(J-1)(J-2)$.*

*Proof*: see Appendix II.

### B. Network with finite backlog and finite number of users

In this case the nodes with empty queue will not contend for medium access. Throughput analysis for this case is carried our by extending the approach of [15] to the case in which the receiver can resolve second order collisions with a certain probability. The beauty of the method in [15] is that it approximates the performance of $J$ coupled queues with $J$ uncoupled $geom/geom/1$ queues, an approximation that simplifies the analysis greatly.

The assumptions in this section follows those in [15], i.e.,

- The arrival rate for each queue in the system is Bernoulli with rate $r$, i.e., the total arrival rate for a system with $J$ users is $rJ$.

- A queue $k$ is active in a time slot if it has one or more packets eligible for transmission, else it is inactive.

- Each active queue $k = 1, ..., J$ contends with a common fixed contention probability $p$.

We further assume that there is an acknowledgement feedback loop, so that the transmitter knows whether the packet that was transmitted was successfully received or has to be re-transmitted. Assuming that the probability that a queue is active in a typical time slot in steady state is $q$, the probability of success for a queue becomes:

$$s(q) = P_0 p \left(1 - qp\right)^{J-1} + \left(P_1 + \frac{P_2}{2}\right) (J-1) p^2 \left(1 - qp\right)^{J-2} \tag{11}$$

where we assumed that in the case of two message transmissions where only one message is successfully received, it could be any of the two messages with equal probability, i.e., we assume all the links to be equivalent.

*1) Active Probability $q$:* Applying Little's Law to the server, we find that $q = \frac{r}{s}$, where $r$ is the arrival rate and $s$ is given by (11). Following the steps from [15], let us define $f(z) = P_0 z \left(1 - z\right)^{J-1} + \left(P_1 + \frac{P_2}{2}\right) (J-1) z^2 \left(1 - z\right)$ and $f^{max} = f(p^*)$, where $p^*$ is the maximizer for $f(z)$. Based on Appendix I $p^* = \frac{2P_0}{2P_0 + a + \sqrt{a^2 + 4P_0 b}}$. The function $f(z)$ corresponds to the success probability of a queue in the system when the queues are unstable and thus always active. In other words, $f(z)$ corresponds to the maximum possible success probability. Since a queue cannot output more packets than the ones that arrive in the queue, we can distinguish between two different modes of operation of the queue as a function of the arrival rate:

For $\underline{r : r > f^{max}}$ the arrival rate in the queue is larger than the maximum possible rate that the packets can exit the queue. In this case the queue is always active (i.e., $q = 1$), and it's success probability is simply $f(p)$, with p the contention probability.

For $\underline{r : r < f^{max}}$ the stability of the queue depends on the contention probability, since the physical layer can support a departure rate greater than the arrival rate. But the queue is not stable for all possible contention probabilities. The equation $f(p) = r$ in this case has two real solutions, let these be $p^{min}$ and $p^{max}$. For $p < p^{min}$ and $p > p^{max}$, the queue is unstable and the active probability $q = 1$. This instability is due to either a very conservative choice of contention probability (for the $p < p^{min}$ case) or a very aggressive one (for the case of $p > p^{max}$). On the other hand, for $p \in \left[p^{min}, p^{max}\right]$, the queue becomes stable (active probability $q < 1$), as in this region of operation the physical layer can support a greater departure rate than $r$. Since a queue cannot output more packets than the ones arriving in the queue, we conclude that the departure rate in this region equals the arrival rate in the queue. In order to calculate the active probability in this region of operation we can simply solve the equation $f(qp) = r$. It is straight forward to see that the two solutions of the equation are $qp = p^{min}$ and $qp = p^{max}$. Solving for $q$, we get that $q = \frac{p^{min}}{p}$ and $q = \frac{p^{max}}{p}$. Since $p \in \left[ p^{min}, p^{max} \right]$ and $q \in [0, 1]$, the only possible solution is $q = \frac{p^{min}}{p}$.

Summarizing, the active probability of each node equals

$$q = \begin{cases} \frac{p^{min}}{p}, & r < f^{max} \text{ and } p \in \left[ p^{min}, p^{max} \right] \\ 1, & \text{otherwise} \end{cases} \tag{12}$$

The equations $f(p) = r$ and $f(qp) = r$ can efficiently be solved for $p$ and $q$ using any numerical method, for example Newton's method.

*2) Approximate Throughput:* The throughput of $J$ independent queues, using $q$, $p^{min}$, $p^{max}$ and $f^{max}$ is

$$\tau = \begin{cases} Jr, & r < f^{max} \text{ and } p \in \left[ p^{min}, p^{max} \right] \\ J P_0 p (1-p)^{J-1} + J \left( P_1 + \frac{P_2}{2} \right) (J-1) p^2 (1-p)^{J-2}, & \text{otherwise} \end{cases} \tag{13}$$

The system throughput (average number of successful transmissions per slot), when queues are stable is limited by the rate at which messages arrive at each of the queues, while in the region where the queues are unstable ($q = 1$), the throughput is limited by the maximum achievable throughput of the physical layer, as in the case of infinite backlog.

*3) Average Total Delay (Queue+Service delays):* For the regions of operation where the active probability $q$ is less than 1 and thus the queue is stable, we can further calculate the total delay a packet will experience from the time it enters the queue, until it is successfully transmitted. As is shown in section IV-B.1, the queue is stable when $r < f^{max}$ and $p \in \left[ p^{min}, p^{max} \right]$. Using the well-known results from queuing theory for the geom/geom/1 queue, the total delay (queuing plus service delay) equals [8]

$$D_{tot} = \frac{1}{s \left( 1 - \frac{r(1-s)}{s(1-r)} \right)}, \quad r < f^{max} \text{ and } p \in \left[ p^{min}, p^{max} \right] \tag{14}$$

where $r(1-s)$ is the "birth probability", $s(1-r)$ is the "death probability" of the queue and $s$ comes from (11), after we calculate the active probability $q$ from eq. 12.

*4) Average Delay in Server:* Since for an geom/geom/1 queue with service rate $s$ the average service delay is $\delta = s^{-1}$, the average service delay of the K independent queues, using $q$, $p^{min}$, $p^{max}$ and $f^{max}$ is

$$\delta = \begin{cases} \frac{q}{r}, & r < f^{max} \text{ and } p \in \left[ p^{min}, p^{max} \right] \\ \left[ P_0 p (1-p)^{J-1} + \left( P_1 + \frac{P_2}{2} \right) (J-1) p^2 (1-p)^{J-2} \right]^{-1}, & \text{otherwise} \end{cases} \tag{15}$$

## V. Testbed setting

The proposed approach was implemented on the WARP testbed [3, 2]. WARP hardware consists of two components: digital baseband processing and analog RF processing. Baseband processing is all done on the main system board which houses a Xilinx Virtex II Pro FPGA for all digital baseband PHY and MAC layer functionality. The main board contains four sets of connectors which provide a set of digital connections to four possible daughtercards that fit onto the main board. These daughtercards perform ADC and DAC functions along with up- and down-conversion to and from the ISM and UNII frequency bands.

In this study we used the non-real time stage of the WARP testbed, which makes use of an API called WARPLab. WARPLab allows all processing and modulation to be done in Matlab, turning the FPGA of WARP into a simple buffer. Matlab can be used to create a set of data, modulate it, apply the designed pulse shaping function, and transfer the data to the radio card. On the receive side, WARPLab allows for data to be processed in Matlab immediately after it has been downconverted by the RFIC on the radio card.

Fig. 2 shows an experimental configuration where a single host computer controls five nodes. The host computer acts as the BS and controls all the nodes in order to provide the correct synchronization between the transmitters and the receiver. The separation between nodes 1 and 2 was about $5$ meters, and the separation between the node 3 and 4, and nodes 4 and 5 was also about $5$ meters. The separation between nodes 1 and 3, and nodes 2 and 5 was about $10$ meters. All the nodes transmitted narrowband signals simultaneously, both using the same carrier frequency $2.447$ GHz (channel N0.8 of 802.11b WiFi channel).

## VI. Details on the SDR implementation

The user packet is structured as shown in Fig. 1. The SDR implementation was carried out in the following steps.

At the transmitter:

- *Paylaod* - The payload contained $414$ bits (32 bits for the user ID and $382$ random bits). Convolutional coding with rate $1/2$ was applied to get $828$ bits. The coded bits were then interleaved. Specifically, the interleaver writes the input sequence in a matrix in row-wise fashion and then reads it in column-wise fashion. Differential quadrature phase shift keying (DQPSK) was used to modulate the data. The IOTA pulse shape waveform was used for transmission.

- *Pilots* - A 32 bit m-sequence was added at the beginning as a sequence of pilots; it was BPSK modulated and RRC pulse shape waveform was used for the transmission of the pilot symbols. A code book of four m-sequences was generated. The code book was kept at the BS and linked with to the user IDs.

- *Sampling rate* - The sampling rate of the board was 40 Msamples/second and 32 samples/symbol were taken, yielding data rate of 1.25 Msymbols/second.

- *Introducing user delays* - A random number of zero samples, chosen uniformly in $[0, 32]$, was added in the beginning of the payload.

- *Transmission* - The signal was first up-converted to 5 MHz and sent to the transmission buffer. The board used channel 8 of the IEEE 802.11 standard to transmit the signal, i.e., the carrier frequency was 2.44GHz.

At the receiver:

- *Synchronization* - The signal was read from the receiver buffer where it was already down-converted to 5 MHz. Subsequently it was down-converted to baseband. All entries of the code book were used to perform correlations with the header of the received signal. The entry which gave the largest correlation peak was chosen to indicate who the corresponding user was. For the following discussion, suppose that this is user $u_1$ ($u_1$ could be any of the users present in the system).

  The delay and channel coefficient of user $u_1$ was estimated based on the location and value of the peak, respectively. The chosen m-sequence was deflated from the pilot portion of the received signal. All entries of the code book were used to perform correlation with the pilot portion of the deflated signal. The entry that gave the largest correlation peak indicated the second user, $u_2$, and the corresponding delay and channel coefficient was estimated.

  Note that if there was only one user, the remaining signal after deflation would be just noise. The same would hold in the case that the received signal contained more that 2 users.

- *Symbol recovery* - The received signal was up-sampled by 4, with sampling points occurred at $[57, 62, 67, 72]$ taps of each received pulse. The resulting 4 polyphase components were input to the JADE algorithm for source separation. The output of the JADE algorithm was input to a PLL. The results were 4 sequences, i.e., $s_{u_1}(i), s_{u_1}(i+1), s_{u_2}(i), s_{u_2}(i+1)$, within phase and delay ambiguities. Let $s_u(.)$ be the strongest signal, i.e., the sequence that has the smallest variance around the known constellation. The symbols corresponding to $s_u(.)$ were demodulated. The use of DQPSK modulation allowed for removal of phase ambiguity. The demodulated output was passed through a de-interleaver and decoder, to get 414 decoded bits. The result could be either $s_u(i)$ or $s_u(i+1)$. If we incorrectly misinterpreted $s_u(i)$ for $s_u(i+1)$, the de-interleaver would give a meaningless output. We used the user ID part in the beginning of the decoded output to do correlation with the corresponding entry of the user ID book in order to determine whether the recovered signal was $s_u(i)$ or $s_u(i+1)$, and also whether the recorded signal corresponded to user $u_1$ or user $u_2$.

Note that although we use correlation with the user IDs to determine the user, we cannot use this information to estimate the channel. This is because the received packet is interleaved and coded, thus the beginning part of the frame is a random sequence until decoding.

Next, we used the detected $s_u(i)$ to obtain the corresponding channel estimate using cross-correlation with the IOTA pulse. Note that we already had obtained a channel estimate for that user during the synchronization step. However, the estimate obtained based on the recovered symbols would be more robust as it is based on 414 symbols; the estimate obtained during the synchronization step was based on 32 symbols. Finally, we deflated the corresponding signal from the received mixture. In the testbed experiment, we assumed that we knew the collision order in each transmission. If the transmission was collision free, we would stop after the first round of detection. In a real case, where the receiver is not aware of the collision order, we would proceed with SIC. The collision free signal would yield a meaningless remainder after SIC and that signal would not pass a checksum verification.

## VII. Testbed measurements

### A. A two-user system

In this experiments, nodes 3 and 5 are the two transmitters, and node 2 is the BS. For each time slot both nodes transmit with probability 1.

*1) BER comparison:* In this section we show the testbed performance of the ALOHA-CR using blind source separation followed by SIC, described in Section III-B (denoted in the figures as *blind*), ALOHA-CR using training based source separation followed by SIC, described in Section III-D, (denoted in the figures as *training*), and ALOHA-CR using SIC only, described in Section III-B, (denoted in the figures as *SIC*). The blind source separation algorithm used was the JADE method [5], which was downloaded from: http://www.tsi.enst.fr/~cardoso/guidesepsou.html.

We first consider the raw BER performance (BER before decoding) of the proposed scheme. In this scenario, the location of the transmitter and receiver, and the antenna gains are fixed. By varying the amplitude of the input signal we can look at the BER performance at different SNR levels. For each SNR level, 600 packets were transmitted from the sender to the BS. Since the indoor wireless channel is time varying in both phase and amplitude, the received SNR of the two users varies between transmissions. The SNR difference of user 1 and user 2 was within $3dB$; $94\%$ of them were within $1dB$.

For comparison purposes only, in this BER evaluation we only include the delay differences in the range $[T_s/2 - T_s/8, T_s/2 + T_s/8]$. When the delay differences are smaller, all methods yield high BER and their performance is the same. The BER performance of the blind approach as captured by the testbed is shown in Fig. 3. One can see that the proposed blind separation scheme works very well. The BER approaches $10^{-3}$ at an SNR of about $20dB$. The performance of the training method is also shown in Fig. 3. One can see for the testbed measurements results there is about $3dB$ performance advantage of the blind over the training method when the SNR is smaller than $20dB$, and this advantage further increases at higher SNR level. The inferior performance of the training method is due to the sensitivity of least-squares at low SNR. Moreover, in the testbed measurements there is distortion of the pulse shape due to the antenna, drifting of the sampling point, and error in the channel coefficient estimates, which also results in degradation of BER performance. The performance of the SIC method is also included in Fig. 3. We can see that there is an error floor which does not decrease with increasing SNR. This is due to the fact that when we attempt to detect the first user, we treat the other user as interference.

Computer simulations were also conducted to produce the BER for this case. In the simulation the channel coefficients, $a_k$, $k = 1, 2$, were taken to have amplitude one and random phase. It was assumed that the channel remains the same within each block. The delays and CFOs were set equal to the values observed during the testbed experiments. The estimation results were averaged over $100$ independent channels, and $10$ Monte-Carlo runs for each channel. In Fig. 3 one can see that there is only $1dB$ gap between the testbed measurements and the computer simulations.

*2) Throughput comparison:* The throughput performance of the three methods is given in Fig. 4. In this figure all received packets are taken into account. We assume that any error in the decoding output results in failure of transmission. The throughput was computed as the number of successful delivered packet per time slot. We can see that, as expected, the blind separation method gives the highest throughput. The throughput of SIC is bounded by $0.3$, and does not increase with increasing SNR level.

Comparison of blind and training methods in terms of throughput fairness between the users is given in Fig. 5. We can see that in the high SNR region ($SNR \geq 20dB$), the throughput of the two users is almost the same; in the low SNR region, the throughput of user 1 is a little bit lower than user 2, which is caused by the SIC. As we first detect and deflate user 1 from the mixture of received signal, the signal to interference and noise ratio (SINR) of user 1 is lower than that when we detect user 2. This SINR difference will be in favor of user 2 in the low SNR case. In the high SNR region, user 1 can be detected very well even in the presence of user 2. Then the throughput of these two users are almost the same.

*3) Throughput vs different shift scenarios:* In this section we demonstrate the advantage of intentional delays. In the previous experiment, we introduce random delay to all users. In this experiment we look at the performance without any intentional delays. The results are shown in Fig. 6. It is clear that if we do not introduce any delays and let nodes transmit freely, the throughput is quite low. This is because the naturally occurring delays are small, thus affecting the condition number of $\mathbf{A}$ and resulting in high BER.

### B. Buffered Slotted Aloha Measurements

For this set of measurements, we employed 5 SDR nodes, as depicted in Fig. 2. Node 5 had the role of the BS and all the other nodes were trying to communicate with it. The transmitted messages consisted of random bits and the length was $414$ before coding. Upon reception, the message was decoded and the transmission was considered successful if there were no bits in error. Each node had an independent Bernoulli arrival process of rate $r$, which resulted in a system arrival rate of $4r$. For the measurements, $r$ took the values of $\frac{1}{2}, \frac{1}{4}, \frac{1}{8}, \frac{1}{16}$ and $\frac{1}{32}$, so as to measure the system performance in various loads. The contention probability for each of the above arrival rates took values in $[0.05, 0.95]$ with steps of $0.05$.

In order to gather meaningful data, we had to make sure that the system was at steady state. Since the actual transmission and reception operations were time-consuming, the measurement process was performed in two steps. For each arrival rate and contention probability the system started at the empty state(all queues empty) and for the first 100,000 slots we were not performing any actual transmissions, but rather decided on the outcome of each slot based on the values of $P_0$, $P_1$ and $P_2$ that were measured off-line for this topology (namely, $P_0$ was measured to be 0.998, $P_1$ was 0.965 and $P_2$ was found to be 0.009) and the number of contending stations. To be more specific, when no transmitters were trying to transmit, the slot was considered empty. When only one transmitter was trying to access the medium, its queue would decrease by one with probability $P_0$. When two transmitters were contending for the medium there was probability $P_1$ of both transmissions being successful, i.e., both of the queues would decrease their size by one with probability $P_1$, and with probability $P_2$ only one of the transmitters would decrease its queue. In the later case, where only one of the two nodes was successful, the successful transmission was assigned on either of the two contending transmitters with equal probability (i.e., probability of 0.5 for each). If more than 2 transmitters were trying to access the medium, a collision was declared and no queue would decrease its size.

After these initial 100,000 slots, there were 3,000 extra slots during which actual transmissions were employed and data for service delay, total delay, throughput and active probability was gathered. For each slot, the outcome was determined by the receiver, depending on how many messages it was able to receive without any errors. Any message that was successfully received, was removed from the corresponding queue, but in case of errors the message had to remain in the queue and be re-transmitted until being successfully received.

The data that was gathered is plotted against the analytically calculated values that were determined as is described in section IV-B. In Fig. 7 the results for the active probability appear, in Fig. 8 we plot the measured and numerically calculated throughput and in Figs. 9 and 10 the measured and analytical results for the total and service delays respectively are plotted. From the plot of the total delay, the lines that correspond to arrival rate of 1/2 do not appear, since in this case the queues are unstable for all possible contention probabilities and thus the total delay goes to infinity. For the active probability and throughput we see that there is almost a perfect match between the measured and the analytically predicted quantities, using the independent queue approach, while for the delay quantities the match is still pretty good, even though it is not as good as for the throughput and active probability. Comparing the system with the conventional buffered slotted Aloha, looking for example at the results from [15], were no collisions can be resolved, we can see that the achieved throughput for ALOHA-CR is more than doubled, the service and total delays are considerably less and further, the system is stable for a much greater span of arrival rates and contention probabilities.

## VIII. CONCLUSION

In this paper, we have proposed ALOHA-CR, which is a novel cross-layer scheme for high throughput wireless communications in a cellular scenario. This scheme can resolve second order collisions in the network without requiring retransmissions. We have described in detail the physical and MAC layers of the proposed scheme and derived analytical expressions to predict its performance. Further, the proposed scheme was implemented in a 5 node SDR system and its measured performance showed very good agreement with the analytical derived results. The conducted measurements show that such a system can achieve more than twice the throughput of a conventional slotted Aloha scheme, while maintaining stability for a much wider range of arrival rates and contention probabilities. This indicates that Aloha-CR might be an excellent option for system deployments that can afford some extra complexity on the access point, while requiring low transmitter complexity (compared to other collision resolution schemes) to meet power or pricing requirements.

## Appendix I

### Proof of the Proposition 1

*Proof*:

Let $f(x)$ be the pdf of the relative delay $\tau$ between the two users. The probability that the collision is not resolvable is:

$$P_c = \sum_{n=-\infty}^{\infty} \int_{-\Delta/2+nT_s}^{\Delta/2+nT_s} f(x)dx, \tag{16}$$

where $\Delta$ is some number smaller that $T$ representing the smallest distance between the peaks of the two users that still allows the users to be resolved.

In (16) $n$ runs from $-\infty$ to $\infty$, because when the relative delay $\tau$ is increased by by $nT_s$, $n \in \mathbb{Z}$, the channel channel matrix $\mathbf{A}$ remains the same.

Because the intentional delays are uniformly distributed in $[0, T]$, the pdf of $\alpha$ is :

$$f_\alpha(x) = \begin{cases} 1/T - |x|/T^2 & \text{if } |x| \leq T \\ 0 & \text{otherwise} \end{cases} \tag{17}$$

Since $\tau = \alpha + \delta$, The PDF of $\tau$ is:

$$\begin{aligned} f(x) &= \int_{-T}^{T} f_\alpha(v)f_\delta(x-v)dv \\ &= \int_0^T (\frac{1}{T} - \frac{v}{T^2})f_\delta(x-v)dv + \int_{-T}^0 (\frac{1}{T} + \frac{v}{T^2})f_\delta(x-v)dv \end{aligned} \tag{18}$$

Substituting (18) into (16), the probability of collision can be represented as:

$$P_c = \int_C \int_0^T (\frac{1}{T} - \frac{v}{T^2})f_\delta(x-v)dvdx + \int_C \int_{-T}^0 (\frac{1}{T} + \frac{v}{T^2})f_\delta(x-v)dvdx \tag{19}$$

where $\int_C dx = \sum_{n=-\infty}^{\infty} \int_{-\Delta/2+nT_s}^{\Delta/2+nT_s} dx$. Now $P_c$ is a function of $T$. Taking the first order derivative of $P_c$ with respect to $T$, we have

$$\frac{dP_c}{dT} = \underbrace{\frac{1}{T^2} \int_C \int_0^T (\frac{2v}{T} - 1)f_\delta(x-v)dvdx}_{\Xi_1} + \underbrace{\frac{1}{T^2} \int_C \int_0^T (\frac{2v}{T} - 1)f_\delta(x+v)dvdx}_{\Xi_2}. \tag{20}$$

Next we will show $\frac{dP_c}{dT}|_{T=T_s} = 0$. Defining $\Phi(x) = \int_{-\infty}^x f_\delta(v)dv$ and $\int_a^b f_\delta(v)dv = \Phi(a) - \Phi(b)$, we get

$$\Xi_1|_{T=T_s} = \frac{1}{T_s^2} \int_0^{T_s} (\frac{2v}{T_s} - 1) \sum_{n=-\infty}^{\infty} \Phi(\Delta/2 + nT_s - v) - \Phi(-\Delta/2 + nT_s - v) dv$$

$$= \frac{1}{T_s^2} \int_{-T_s/2}^{T_s/2} \frac{2u}{T_s} \underbrace{\sum_{n=-\infty}^{\infty} \Phi(\Delta/2 + nT_s - u - T_s/2) - \Phi(-\Delta/2 + nT_s - u - T_s/2)}_{\phi(\mathbf{u})} du \quad (21)$$

where $u = v - T_s/2$. As $2u/T_s$ is an odd function in $[-T_s/2, T_s/2]$, if $\phi(u)$ is an even function of $u$, then $\Xi_1|_{T=T_s} = 0$. Indeed, since $\Phi(x) = 1 - \Phi(-x)$, it can be easily seen that $\phi(-u) = \phi(u)$. Similarly we can show that $\Xi_2|_{T=T_s} = 0$. Thus $\frac{dP_c}{dT}|_{T=T_s} = 0$.

Next we show that $\frac{d^2P_c}{dT^2}|_{T=T_s} > 0$.

$$\frac{d^2P_c}{dT^2} = \frac{d\Xi_1}{dT} + \frac{d\Xi_2}{dT} \quad (22)$$

$$\frac{d\Xi_1}{dT} = \frac{-6}{T^4} \int_C \int_0^T v f_\delta(x-v) dv dx + \frac{2}{T^3} \int_C \int_0^T f_\delta(x-v) dv dx + \frac{1}{T^2} \int_C f_\delta(x-T) dx$$

$$= -\frac{2}{T} \Xi_1 + \frac{1}{T^2} \int_C f_\delta(x) dx - \frac{2}{T^4} \int_C \int_0^T v f_\delta(x-v) dv dx, \quad (23)$$

where $\int_C f_\delta(x) dx = \int_C f_\delta(x-T) dx$. Let us assume that $\int_C f_\delta(x) dx > \frac{1}{T_s} \int_0^{T_s} \int_C f_\delta(x-v) dx dv$. This means that the probability for non resolvable collisions when we do not introduce any intentional random delays to any user is larger that that when we only introduce any intentional random delays to one of the two users involved in the collision. The is intuitively correct, and was further confirmed in our testbed.

By applying $\Xi_1|_{T=T_s} = 0$, we get

$$\frac{d\Xi_1}{dT}|_{T=T_s} > \frac{1}{T_s^3} \left[ \int_0^{T_s} \int_C f_\delta(x-v) dx dv - \frac{2}{T_s} \int_C \int_0^{T_s} v f_\delta(x-v) dv dx \right]$$

$$= -\frac{\Xi_1|_{T=T_s}}{T_s} = 0. \quad (24)$$

Similarly, we can prove that $\frac{d\Xi_2}{dT}|_{T=T_s} > 0$, which leads to $\frac{d^2P_c}{dT^2}|_{T=T_s} > 0$. Thus we has shown that if we assign an intentional delay $\tilde{\tau}_k$ to each user that is uniformly distributed in $[0, T_s]$, the collision probability achieves a local minimum value.

## Appendix II

## Proof of the Proposition 2

Recall that for a single transmission the probability of successful reception is $P_0$; for order two collision the probability of receiving correctly both of the transmission messages is $P_1$ and the probability of successfully receiving only one of the two transmitting messages is $P_2$. The throughput of this network is :

$$
\begin{aligned}
C(p) &= P_0 J p (1-p)^{J-1} + (2P_1 + P_2) \binom{J}{2} p^2 (1-p)^{J-2} \\
&= P_0 J p (1-p)^{J-1} + \underbrace{\left(P_1 + \frac{P_2}{2}\right)}_{P'} J(J-1) p^2 (1-p)^{J-2}
\end{aligned}
\tag{25}
$$

Let us find the value of $p$ that maximizes throughput. Taking the derivative of $C(p)$ with respect to $p$, we get

$$
\frac{dC(p)}{dp} = J(1-p)^{J-3} [P_0 (1-p)^2 - \underbrace{(P_0 - 2P')(J-1)}_{a} p(1-p) - \underbrace{P'(J-1)(J-2)}_{b} p^2].
\tag{26}
$$

Forcing $dC(p)/dp = 0$, besides a trivial solution at $p = 1$, we get two zeros at

$$
\begin{cases}
p_1^* = \frac{2P_0 + a - \sqrt{a^2 + 4P_0 b}}{2(P_0 + a - b)} = \frac{2P_0}{2P_0 + a + \sqrt{a^2 + 4P_0 b}} \\
p_2^* = \frac{2P_0 + a + \sqrt{a^2 + 4P_0 b}}{2(P_0 + a - b)} = \frac{2P_0}{2P_0 + a - \sqrt{a^2 + 4P_0 b}}.
\end{cases}
\tag{27}
$$

Because $b > 0$, then $a - \sqrt{a^2 + 4P_0 b} < 0$. The possible range for $p_2^*$ is $(-\infty, 0)$ or $(1, \infty)$, which violates the requirement that $0 < p < 1$. Hence only $p_1^*$ is the valid solution. Moreover it is easy to see that when $0 < p < p_1^*$, $dC(p)/dp > 0$, while $p_1^* < p < 1$, $dC(p)/dp < 0$. Thus $C(p)$ is maximized when $p = p_1^*$. As $a$ and $b$ are related to $J$, defining $\eta(J) = 2P_0 + a + \sqrt{a^2 + 4P_0 b}$, we get $p_1^* = 2P_0/\eta(J)$,

$$
\lim_{J \to \infty} \frac{J}{\eta(J)} = \frac{1}{P_0 - 2P' + \sqrt{P_0^2 + 4P'^2}},
\tag{28}
$$

and

$$
\begin{aligned}
\lim_{J \to \infty} \left(1 - \frac{2P_0}{\eta(J)}\right)^J &= \lim_{J \to \infty} \left(1 - \frac{2P_0}{\eta(J)}\right)^{\frac{\eta(J)}{2P_0} \lim_{J \to \infty} \frac{2P_0 J}{\eta(J)}} \\
&= e^{\frac{-2P_0}{P_0 - 2P' + \sqrt{P_0^2 + 4P'^2}}}
\end{aligned}
\tag{29}
$$

Substituting $p_1^* = 2P_0/\eta(J)$ into (25), and based on (28) and (29), we have

$$
\lim_{K \to \infty} C(p_1^*) = \frac{2P_0^2}{P_0 - 2P' + \sqrt{P_0^2 + 4P'^2}} \left(1 + \frac{2P'}{P_0 - 2P' + \sqrt{P_0^2 + 4P'^2}}\right) e^{\frac{-2P_0}{P_0 - 2P' + \sqrt{P_0^2 + 4P'^2}}},
\tag{30}
$$

which gives us the asymptotical throughput when then number of users increases.

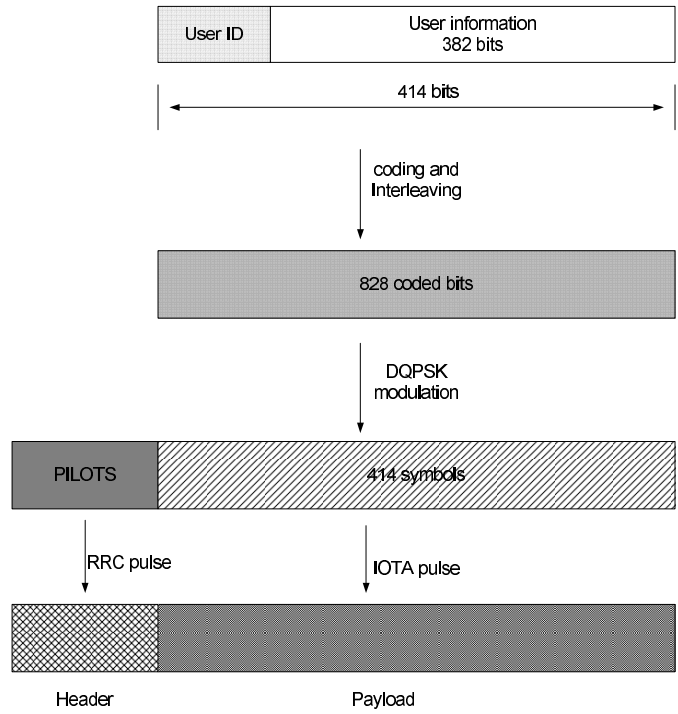

Fig. 1.   Signal Structure

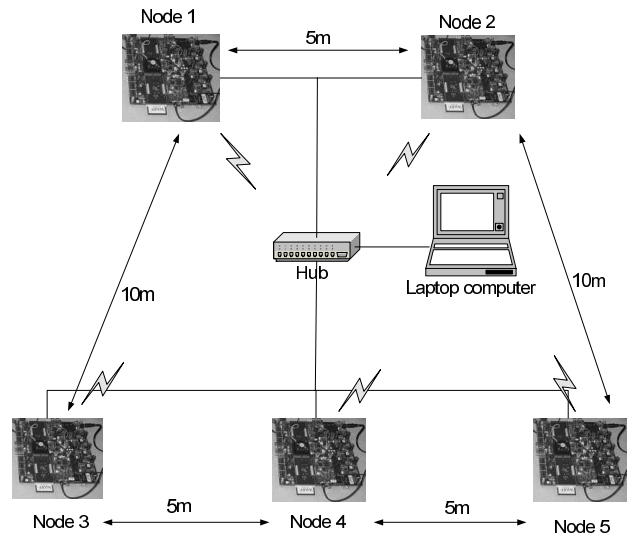

Fig. 2. Deployment of experiment

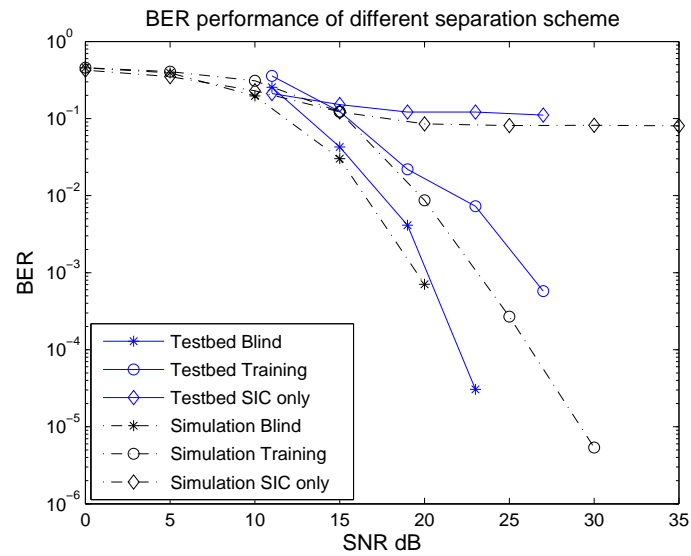

Fig. 3. BER performance of different separation scheme

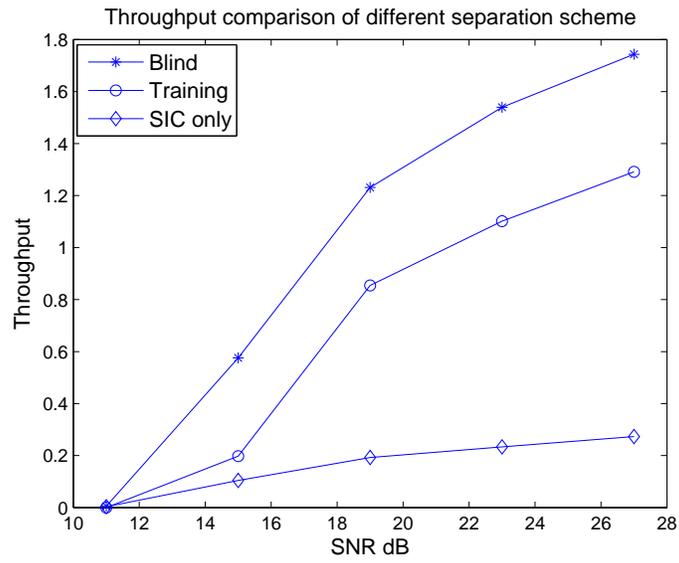

Fig. 4.    Throughput performance of different separation scheme

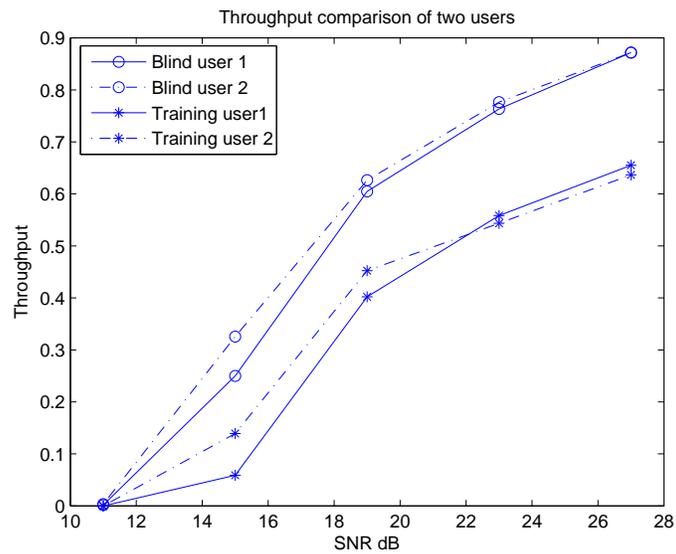

Fig. 5.    Throughput comparison of different users

                                                    

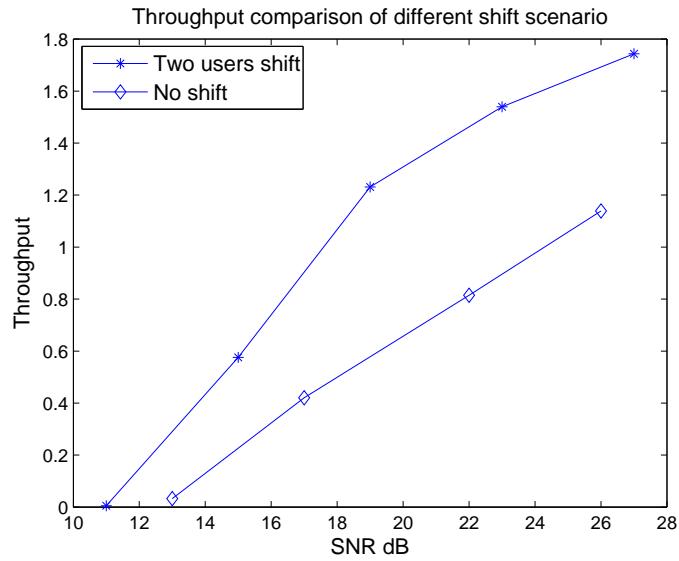

Fig. 6. Throughput comparison of different shift scenarios

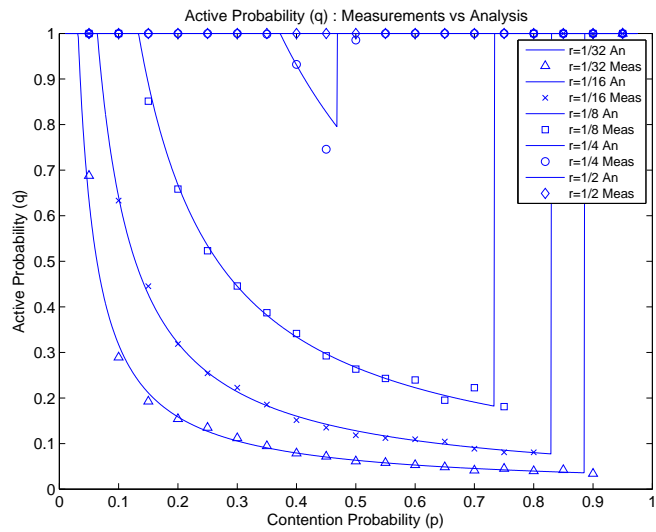

Fig. 7. Active Probability vs contention probability of Aloha-CR for different arrival rates (4 users)

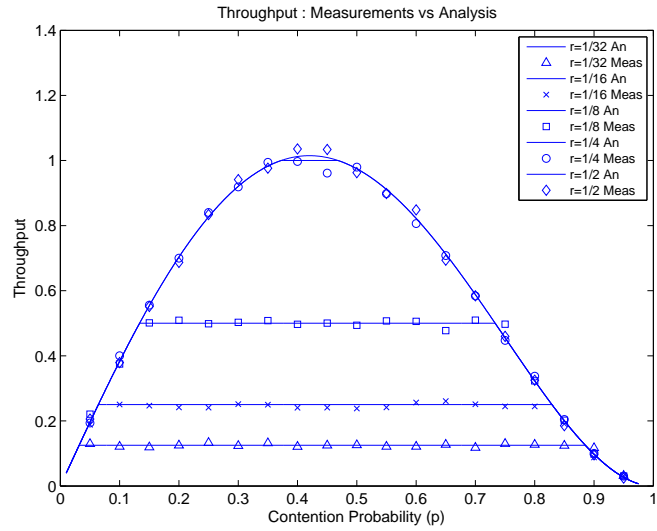

Fig. 8.   Throughput of Aloha-CR vs contention probability for different arrival rates (4 users)

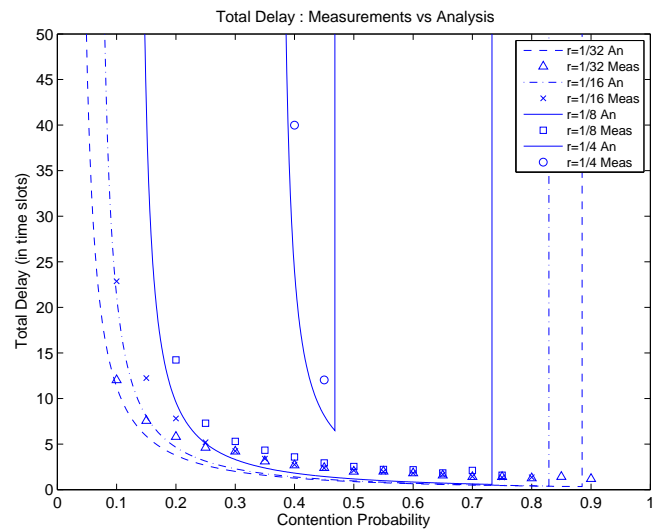

Fig. 9.   Total delay of Aloha-CR vs contention probability for different arrival rates (4 users)

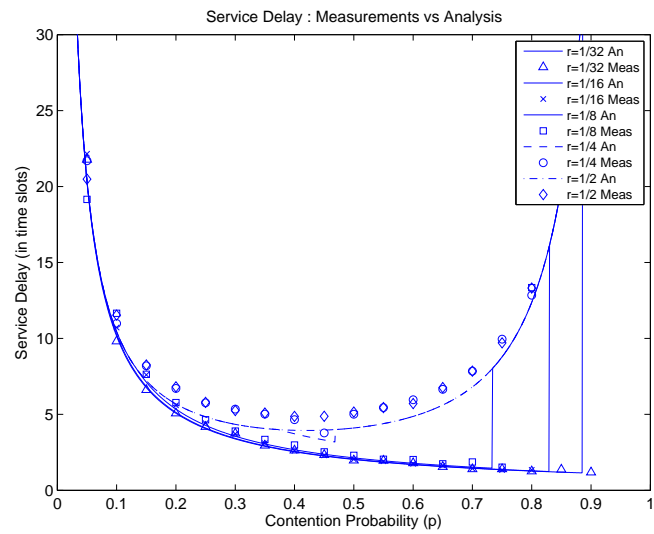

Fig. 10. Service delay of Aloha-CR vs contention probability for different arrival rates (4 users)